\documentclass[prb,twocolumn,superscriptaddress]{revtex4-2}
\usepackage{amsfonts}
\usepackage{amssymb}
\usepackage{amsmath}
\usepackage{graphicx}
\usepackage{dcolumn}
\usepackage{bm}
\usepackage{units}
\usepackage{soul}
\usepackage{multirow}
\usepackage{CJKutf8}
\usepackage{subfigure}
\usepackage{color}%
\usepackage{url}
\usepackage[colorlinks,linkcolor=blue,anchorcolor=blue,citecolor=blue,urlcolor=blue]{hyperref}
\usepackage{array}
\usepackage{ulem}
\begin{document}
\title{A theory for anisotropic magnetoresistance in materials with two vector order parameters}
\author{X. R. Wang}
\email{Chin. Phys. Lett. 39, 027301 (2022)}
\affiliation{Department of Physics, The Hong Kong University of Science and Technology, 
	Clear Water Bay, Kowloon, Hong Kong, China}
\affiliation{HKUST Shenzhen Research Institute, Shenzhen, 518057, China}
	
\begin{abstract}
Anisotropic magnetoresistance (AMR) and related planar Hall 
resistance (PHR) are ubiquitous phenomena of magnetic materials. 
Although the universal angular dependences of AMR and PHR in magnetic    
polycrystalline materials with one order parameter are well known, no similar 
universal relation for other class of magnetic materials are known to date. 
Here I present a general theory of galvanomagnetic effects in magnetic materials 
with two vector order parameters, such as magnetic single crystals with a dominated 
crystalline axis or polycrystalline non-collinear ferrimagnetic materials. 
It is shown that AMR and PHR have a universal angular dependence. In general, both 
longitudinal and transverse resistivity are non-reciprocal in the absence of 
inversion symmetry: Resistivity takes different value when the current is reversed.
Different from simple magnetic polycrystalline materials where AMR and PHR have 
the same magnitude, and $\pi/4$ out of phase, the magnitude of AMR and PHR of 
materials with two vector order parameters are not the same in general, and the 
phase difference is not $\pi/4$. Instead of $\pi$ periodicity of the usual AMR 
and PHR, the periodicities of materials with two order parameters are $2\pi$. 
\end{abstract}
	
\maketitle
Anisotropic magnetoresistance (AMR) of magnetic materials is a well-known 
phenomenon dated back to 1856 when Lord Kelvin \cite{Kelvin}, then known as 
William Thomson, found that electrical resistances of a piece of nickel and iron 
is different when a current is parallel and perpendicular to the magnetization.
This phenomenon and related planar Hall resistance (PHR) are technologically 
important in magnetic sensors and data storage and retrieval \cite{book1,book2}. 
AMR and PHR have constantly attracted much attention with continuously 
improved understanding since their discoveries \cite{CFJ,Tsunoda1,Tsunoda2,
McGuire1,McGuire2,Ohandley,smit,gerrit,Wisniewski,sin-cry-9,sin-cry-10}. 
Phenomenologically and logically, spin-orbit interaction and s-d scatterings 
must play essential roles in the AMR, PHR, and extraordinary galvanomagnetic 
effects \cite{book1,book2} in general because moving electrons ``see" 
the magnetization (spins). Unfortunately, the exact origins of AMR and PHR 
are not clear to date despite of those progress.

AMR can, in principle, occur in all magnetic materials, but is more notable 
for good conducting materials such that magnetization dependent band structure 
and spin dependent scatterings contribute significantly to their resistances. 
It is typically a few per cent for metallic magnetic materials such as iron 
and nickel, and much less for amorphous materials whose resistances are mainly 
from other spin-independent scattering. However, there are also reports that 
AMR could be more than $50\%$ in some single crystals \cite{Wisniewski}. 
To date, the only known universal behaviour of AMR is its angular dependence in 
magnetic polycrystals when the only order parameter is the total magnetization. 
The anisotropic resistivity (or resistance) $\Delta\rho(\alpha)$ follows 
$\Delta\rho(\alpha)=\Delta\rho_0\cos^2\alpha$ where $\alpha$ is the angle 
between the magnetization and current. 
The widely known theory for this universal law is the so-called two 
current model proposed by Campbell, Fert and Jaoul \cite{CFJ,book1,book2}. 
The theory and its extensions \cite{Tsunoda1,Tsunoda2} assume that current 
can be divided into the contributions from spin-up and spin-down electrons. 
Without s-d scatterings, two currents are independent from each other, 
magnetoresistance is determined by the shunted current. In the presence of 
s-d scatterings, two currents are partially mixed and the lift of shunting 
effect depends on the angle $\alpha$ between the magnetization and current. 
The theory leads to an approximate $\cos^2\alpha$-formula for AMR under certain 
assumptions and limits, in contrast to the exact $\cos^2\alpha$-law observed 
in many polycrystalline magnets. Furthermore, these theories require the exact 
angular distribution of atomic d-orbits for explaining the $\cos^2\alpha$-law, in 
contrast to the fact that d-electron wavefunctions in a polycrystal must deviate 
from their atomic counterparts at sub-nanometer scale due to the crystal fields. 
In one word, these theories cannot account the exact universal angular dependences 
of AMR observed in many polycrystalline magnets originated from both d-electrons 
and f-electrons. Of course, it should be pointed out that much less-known theories 
based on symmetry argument \cite{Jungwirth} and tensor analysis \cite{yin1} can 
indeed explain the universal AMR and PHR for polycrystalline magnetic materials. 
For the AMR and PHR in single crystals, there are also many studies \cite{sin-cry-1,
sin-cry-2,sin-cry-3,sin-cry-4,sin-cry-5,sin-cry-6,sin-cry-7,sin-cry-8} that 
show complicate behaves with limited understanding \cite{sin-cry-9,sin-cry-10}. 
The observed behaviours cannot be explained by the two-current model \cite
{Tsunoda1,Tsunoda2} or those \cite{Jungwirth,yin1} for polycrystalline materials. 

In this paper, we derive a generic formula for extraordinary galvanomagnetic effects in 
magnetic materials with two vector order parameters. Materials could be polycrystalline 
ferrimagnets with two sublattices, or magnetic single crystals whose electron 
transport is dominated by one direction while the other two directions are the same. 
It could also be the helimagnets in helical states \cite{haitao}. 
Our theory is based on the general requirement that all physical quantities 
must be tensors \cite{Sakurai} and the laws of physics must be in tensor forms. 
Under this requirement, the AMR and PHR have universal forms in such magnetic  
materials that are distinct from the well known AMR and PHR behaviours 
in simple polycrystalline magnetic materials with only one order parameter.  
The magnitude of AMR and PHR are not the same in general, and the phase difference 
is not $\pi/4$. Instead of $\pi$ periodicity in the usual AMR and PHR, the 
periodicities of the AMR and PHR in the new class of materials are $2\pi$. 

Consider an infinite ferromagnetic single crystal with only one dominate 
crystalline axis $\vec n$ and a magnetization $\vec M$, which is not along $\vec n$. 
In the linear response region, the electric field $\vec E$ in response to an applied 
current density $\vec J$ in the crystal must be 
\begin{equation}
\vec E=\tensor{\rho} (\vec M, \vec n) \vec J,
\label{eq1}
\end{equation}
where $\tensor\rho(\vec M, \vec n)$ is a Cartesian tensor of rank 2. 
Although the tensor values depend on microscopic properties of the crystal and 
parameters that defines its thermodynamic state, the tensor form can only come 
from the order parameters that characterize the macroscopic state of the system. 
In the absence of an external magnetic field, $\vec M$ and $\vec n$ are the 
only available vectors that can be used to construct tensor $\tensor\rho$. 
Thus, $\tensor\rho$ should be the linear combination of $\vec M\vec M$, 
$\vec M\vec n$, and $\vec n\vec n$. Each of the three 
Cartesian tensors is not irreducible \cite{Sakurai}, and can be decomposed 
into the direct sum of a scalar, a vector, and a traceless symmetric tensor. 
Thus, from $\vec M$ and $\vec n$, it is possible to construct three vectors 
and three traceless symmetric tensors of ranks 2. They are $\vec M$, $\vec n$, 
$\vec M\times\vec n$, $\vec M \vec M-M^2/3$, $\vec M \vec n+\vec n\vec M-2\vec 
M\cdot\vec n/3$, and $\vec n\vec n-1/3$, where $M$ is the magnitude of 
magnetization $\vec M$. Thus, with these six angular dependent terms together 
with a scaler term, the electric field $\vec E$ induced by $\vec J$, after 
grouping similar terms, must take the following most generic form  
\begin{equation}
\begin{aligned}
&\vec E=\rho_1 \vec J+ \vec J \times (R_1 \vec M + R_2 \vec n) +
A_1(\vec J\cdot \vec M)\vec M +  \\
&A_2(\vec J\cdot \vec n)\vec n+A_3\vec J\times(\vec M\times \vec n)
+A_4(\vec M\cdot \vec n)\vec J,
\end{aligned}
\label{eq2}
\end{equation}
where $\rho_1$ $R_1$, $R_2$, $A_i$ ($i=1,2,3,4$) are parameters that are 
determined by the intrinsic and extrinsic properties of the sample such as the 
band structures and impurity specifics. Of course, these parameters can, in 
principle, depend on the scalers constructed from $\vec M$ and $\vec n$, such 
as $M^2$ and $\vec M\cdot\vec n$. Among them, only $\vec M\cdot\vec n$ can 
introduce the anisotropic effect. Since magnetic interaction are very weak, 
in terms of perturbation, only low powers of $M^2$ and $\vec M\cdot\vec n$ 
contribute mainly to these parameters. The terms depend on the 
relative directions among $\vec J$, $\vec M$, and $\vec n$. 
It may be worthwhile to point out that this is in the same sprite as those 
in the thermodynamics: The behaviour of a given system can be uniquely 
determined by a few parameters. In the current case, $\vec E$ is 
determined by $\vec J$, $\vec M$, and $\vec n$ when other external 
parameters such as temperature and pressure do not vary.

In the case of polycrystalline sample, $\vec n$ is absent, and Eq. (\ref{eq2}) 
reduce to the well known generalized Ohm's law of polycrystalline materials 
\cite{yin1,yin2} with only $\rho_1$, $R_1$ and $A_1$ terms. $R_1$-term and 
$A_1$-term are the usual anomalous Hall effect and AMR and PHR, respectively. 
The longitudinal and transverse resistivity are $\rho_{xx}=\rho_1+A_1M^2\cos^2
\alpha$ and $\rho_{xy}=R_1M_z+\frac{A_1M^2}{2}\sin 2\alpha$, if the $\hat x$ 
direction is defined along $\vec J$ and $\vec M$ is in the $xy-$plane with angle 
$\alpha$ between $\vec M$ and $\vec J$. Obviously, $\rho_1$ is the longitudinal 
resistivity when $\vec J$ is perpendicular to $\vec M$ and $R_1$ is the anomalous 
Hall coefficient. $A_1M^2$ is the amplitude of the conventional AMR and PHR 
that is typically a few percent of $\rho_1$. 

It may be illustrative to consider two special set-ups of $\vec J$ parallel or 
perpendicular to $\vec n$. For $\vec J \parallel \vec n \parallel \hat x$, $\rho_{xx}
\equiv \vec E\cdot \hat x/J=(\rho_1+A_2)+A_1M^2\cos^2\alpha+A_4M\cos\alpha$ and $\rho
_{xy}\equiv \vec E\cdot \hat y/J=R_1M_z+\frac{A_1M^2}{2}\sin 2\alpha+A_3M\sin\alpha$. 
Interestingly and strangely, it predicts a non-reciprocal dc electron transport if 
the system is not invariant under $\vec n\rightarrow -\vec n$ transformation. Namely, 
$\rho_{xx}$ and $\rho_{xy}$ take different values when current $\vec J$ is reversed. 
Phases of AMR and PHR do not differ by $\pi/4$, and their amplitudes are not the same.
If the system is invariant under $\vec n\rightarrow -\vec n$ transformation, 
then $A_3$ and $A_4$ must be odd function of $\vec M\cdot \vec n=M\cos\alpha$. 
Assume only leading order of $A_3=C_3M\cos\alpha$ and $A_4=C_4M\cos\alpha$ exist, 
then $\rho_{xx}=(\rho_1+A_2)+(A_1+C_4)M^2\cos^2\alpha$ and $\rho_{xy}=R_1M_z+
\frac{A_1+C_3}{2}M^2\sin 2\alpha$. AMR and PHR follow the conventional 
$\cos^2\alpha$ and $\sin 2\alpha$ laws and have the same phase lag but with 
different amplitude, in general. 

For $\vec J \parallel \hat x \perp \vec n \parallel \hat z$, AMR behaves 
differently for $\vec M$ varying in the $xy$- and $yz-$, and $zx-$planes. 
$\rho_{xx}=\rho_1+A_1M^2\cos^2\alpha$ for $\vec M$ in the $xy-$plane and $\alpha$ 
being the angle between $\vec M$ and $\vec J$, $\rho_{xx}=\rho_1+A_4M\cos\beta$ 
for $\vec M$ in the $yz-$plane and $\beta$ being the angle between $\vec M$ and 
$\vec n$, and $\rho_{xx}=\rho_1+A_1M^2\sin^2\gamma+A_4M\cos\gamma$ for $\vec M$ 
in the $zx-$plane and $\gamma$ being the angle between $\vec M$ and $\vec n$.
With the $\vec n\rightarrow -\vec n$ symmetry and with the same reason 
mentioned above, we have $\rho_{xx}(\alpha)=\rho_1+A_1M^2\cos^2\alpha$,
$\rho_{xx}(\beta)=\rho_1+C_4M^2\cos^2\beta$, and $\rho_{xx}(\gamma)=\rho_1+A_1M^2
+(C_4-A_1)M^2\cos^2\gamma$. Clearly, this is very different from the 
conventional AMR that does not have angular dependence for $\vec M$ in the 
$yz$-plane.

Magnetic materials do not respect the time-reversal symmetry because a current reverses 
its direction under the transformation, and so the related magnetization does. Both 
spins and magnetization do not change under an inversion transformation, but a current 
reverses its direction under the transformation. Thus, magnetic materials without inversion 
symmetry are good candidates for observing the non-reciprocal dc electronic transport.   
This can happen for chiral magnets with the Dzyaloshinskii-Moriya interactions (DMI). 
Like the ubiquity of spin-orbit interactions for all materials, DMI universally exists in  
all magnetic materials. Thus, their resistivity should be non-reciprocal in principle, 
and the issue is only whether the non-reciprocity is strong enough to be measurable. 
In case that Eq. \ref{eq2} is invariant under the inversion, $\vec x\rightarrow -\vec x$, 
$\rho_1$, $R_1$, $A_1$, and $A_2$ must be even functions of $\vec M\cdot\vec n$, and 
$R_2$, $A_3$, and $A_4$ are odd since both $\vec E$ and $\vec J$ change sign. 
Thus, the non-reciprocity of resistivity disappears in a system with the inversion 
symmetry, and the angular dependence of AMR and PHR is exactly the same as those for 
the polycrystalline magnetic material with only one order parameter. For polycrystalline 
magnetic-film/heavy-metal bilayer such as $Co/Pt$, the system has the in-plane inversion 
symmetry, but without mirror symmetry with respect to the interface. In this case, one 
expects reciprocity of the longitudinal resistivity in the plane and non-reciprocity in 
tunnelling resistance. Of course, one should anticipate difficulty in detecting the 
interfacial effect in metallic systems. 

The approach above may offer a natural explanation to the strange angular dependences 
of $\rho_{xx}$ in body-centred cubic (bcc) $CoFe$ single crystal when $\vec J\parallel 
\hat x$ is along the magnetic crystalline easy-axis [1,1,0] ($\vec n_1$) and the film 
deposited direction is along [0,0,1] (the $\hat z$-direction) and the $\hat y$ axis is 
along another equivalent easy-axis of $[1,\bar 1, 0]$ ($\vec n_2$) \cite{sin-cry-9}. 
The distinct features include identical strong two-fold AMR when $\vec M$ varies in the 
$yz-$ and $zx-$planes, and a weak four-fold AMR when $\vec M$ varies in the $xy-$plane. 
Although $CoFe$ is not a single crystal with only one axis, one can still construct 
relevant terms in $\vec E$ by treating $\vec n_1$ and $\vec n_2$ as equivalent vectors. 
Then $\vec E=\rho_1\vec J+ \vec J \times R_1\vec M+A_1J M_x^2+A_2J\vec n_1+A_4(\vec M
\cdot\vec n_1)\vec J+A_4'(\vec M\cdot \vec n_2)\vec J$, $R_2$- and $A_3$-terms in Eq. 
(\ref{eq2}) are absent because the equivalent vectors $\pm \vec n_i$ ($i=1,2$) cancel 
each other. $A_1$ must be even in $\vec M\cdot \vec n_1$ and $\vec M
\cdot\vec n_2$, and $A_4$ and $A_4'$ must be odd because $\vec E$ should be symmetric 
under $\vec n_i \rightarrow -\vec n_i$ (i=1,2) transformations for bcc $CoFe$. 
If we take $A_1=C_1(\vec M\cdot \vec n_2)^2$, $A_4=C_4\vec M\cdot \vec n_1$ and $A_4'=C_4
\vec M\cdot \vec n_2$, the longitudinal resistivity becomes, $\rho_{xx}=\rho_1+C_1 J M_x
^2M_y^2+C_4M_x^2 +C_4M_y^2$. It gives $\rho_{xx}(\alpha)=\rho_1'+C_1 J M^4\cos 4\alpha$, 
$\rho_{xx}(\beta)=\rho_1+C_4M^2\sin^2\beta$, and $\rho_{xx}(\gamma)=\rho_1+C_4M^2\sin^2
\gamma$, where $\alpha$ is the angle between $\vec M$ and $\vec J (\hat x)$ when 
$\vec M$ varies in the $xy$-plane. $\beta$ and $\gamma$ are the angles between $\vec M$ 
and $\hat z$ when $\vec M$ varies in the $yz$-plane and $zx$-plane, respectively. 
Since $C_1$ is the 4'th order ($M^4$) and $C_4$ is the second order ($M^2$) in 
perturbation, one expects $C_4\gg C_1$. Thus, four-fold AMR of $\rho_{xx}(\alpha)$ is 
much weaker than the two identical two-fold AMR of $\rho_{xx}(\beta)=\rho_{xx}(\gamma)$. 
This is very unusual because the conventional AMR has a strong two-fold $\rho_{xx}(\alpha)$, 
and no or insignificant angular dependence of $\rho_{xx}(\beta)$ because $\vec J \perp \vec M$. 
These unusual angular dependences of $\rho_{xx}$ on orientation of $\vec M$ is exactly 
what was observed in the recent experiment \cite{sin-cry-9}. If $C_1$ and $C_4$ are the 
intrinsic mechanisms of AMR that come from the modification of band by magnetization 
$\vec M$ when it aligns along the crystalline axis [1,1,0] and $[1,\bar 1,0]$, then 
above $\vec E$ expression makes a lot of sense. It may be important to emphasise that 
above analysis works only when the current is along $[1,1,0]$. Furthermore, above 
analysis depends not only on bcc $CoFe$, but also on the assumption that, somehow, only 
two equivalent directions of $[1,1,0]$ and $[1,\bar 1, 0]$, but not others, are important. 
The analysis provides only a possible understanding of the recent surprising experimental 
observations, and more study is needed.    

Magnetic single crystals with only one dominated axis are not very common because a crystal 
has three principle axis by definition although many tetrahedron and hexagonal structures 
are believed to be uniaxial magnets. To test the theory presented above, it may be useful 
to find easily realizable materials where the generalization of the theory applies. Real 
crystals may have other rotational and mirror symmetries that provide extra macroscopic orders. 
There would be many more terms in Eq. \ref{eq2} and above simple universal behaviour disappear. 
Thus, more realistic materials to test the theory may be polycrystalline materials. As clearly 
shown in our derivation, the theory is based on the assumption of two vector order parameters. 
Thus, the theory is applicable to non-collinear ferrimagnetic polycrystals with two sub-latices. 
The magnetizations $\vec M_1$ and $\vec M_2$ of the two sub-lattices are the only available 
order parameters, and we need only to replace $\vec M$ and $\vec n$ in Eq. (\ref{eq2}) by 
$\vec M_1$ and $\vec M_2$. The resulting equation is 
\begin{equation}
\begin{aligned}
&\vec E=\rho_1 \vec J+ \vec J \times (R_1 \vec M_1 + R_2 \vec M_2) +
A_1(\vec J\cdot \vec M_1)\vec M_1 +  \\
&A_2(\vec J\cdot \vec M_2)\vec M_2+A_3\vec J\times (\vec M_1\times\vec M_2)
+A_4(\vec M_1\cdot \vec M_2)\vec J.
\end{aligned}
\label{eq3}
\end{equation}
Of course, $|\vec M_2|=M_2\neq 1$ now.

Consider a special case where $\vec J$, $\vec M_1$, and $\vec M_2$ are in a plane 
with the angles between $\vec J$ and $\vec M_1$ and $\vec M_2$ being $\theta$ and $\phi$.
The longitudinal and transverse resistivity are 
\begin{equation}
\begin{aligned}
&\rho_{xx}=\rho_1+\rho_2\cos^2\theta+\rho_3\cos^2\phi+\rho_4\cos(\phi-\theta)  \\
&\rho_{xy}=\rho_5+\frac{\rho_2}{2}\sin 2\theta+\frac{\rho_3}{2}
\sin 2\phi+\rho_6\sin(\phi-\theta),
\end{aligned}
\label{eq4}
\end{equation}
where $\rho_2\equiv A_1M_1^2$, $\rho_3\equiv A_2M_2^2$, $\rho_4\equiv A_4M_1M_2$,
$\rho_5\equiv R_1M_{1z}+R_2M_{2z}$, and $\rho_6\equiv A_3M_1M_2$. When $\phi-\theta=
180^0$, or collinear ferrimagnetic crystalline, Eq. (\ref{eq4}) returns to the well-known 
AMR and PHR. It is well known that a collinear antiferromagnet undergoes a spin-flop 
transition when a magnetic field along the Neel order parameter is larger than a critical 
value. Thus, the current theory may be best tested in a collinear antiferromagnet.
The fingerprint is the change of angular dependences of AMR and PHR before 
and after spin-flop transition: Before the transition, AMR and PHR have the same 
amplitudes and follow a $\cos^2\theta$-law and $\sin 2\theta$-law respectively. 
After the transition, AMR and PHR are described by Eq. (\ref{eq4}). If the 
antiferromagnet has a strong DMI such that the system does not have inversion symmetry, 
the magnitude of AMR and PHR of the materials with two vector order parameters 
are not the same, as discussed earlier, and the phase difference is not $\pi/4$. 
Instead of $\pi$ periodicity of the AMR and PHR in polycrystalline materials with only 
one order parameter, the periodicities of the AMR and PHR in Eq. (\ref{eq4}) are $2\pi$.

Our predictions are based on the tensor forms of the laws of physics. 
This is the same approach for the Einstein's gravitational law. 
The argument is that the Ricci tensor and metric tensor are the only possible tensors 
of rank 2 out of the metric tensor while Newtonian gravitation-law was identified 
as an equation between one metric tensor component and the energy \cite{Carmeli}. 
Thus, a proper linear combination of Ricci tensor and Ricci curvature 
(scaler) multiplying metric tensor is equal to the energy-momentum tensor.
The method was also used in condensed matter physics for Ohm's laws in ferromagnet 
\cite{yin2} and for the anomalous spin-Hall effects (ASHEs) and anomalous inverse 
spin-Hall effects (AISHEs) \cite{xrw,yin3}, which is also called magnetization 
dependent spin-Hall and inverse spin-Hall effects. Interestingly, our analysis 
predicts that $\rho_{xx}(\theta=180^0)\neq \rho_{xx}(\theta=0^0)$ in general. 
However, $\rho_{xx}(\theta=180^0)=\rho_{xx}(\theta=0^0)$ holds in a material 
if it has the inversion symmetry. This is because of $\vec E(-\vec J)=-\vec 
E(\vec J)$ under the symmetry. The AMR and PHR in such a material behave the 
same as those in a polycrystal with only one order parameter. 

It should be pointed out that irreducible tensor decomposition in our derivation 
of Eq. (\ref{eq2}) is important because different irreducible tensors vary with 
other parameters independently although they are from the same Cartesian tensors. 
For systems with two vector order parameters, $\rho_1$, $R_i (i=1,2)$, and $A_i (i=1,
2,3,4)$ are the seven independent material parameters in the galvanomagnetic effects. 
The general expressions of Eqs. (\ref{eq2}) and (\ref{eq3}) do not distinguish 
an intrinsic mechanism such as the band structure contribution from an extrinsic 
mechanism such as spin-dependent electron scattering due to defects and phonons. 
Our predictions were not derived from a microscopic Hamiltonian, and the theory  
does not provide actual values of the seven parameters. 
In principle, their values can be computed from a given microscopic model using 
quantum mechanics. Such a microscopic theory is surely important and necessary 
although it is foreseeable not easy because of too many process in real materials. 

In summary, a generic galvanomagnetic effect in magnetic materials whose 
thermodynamic states can be described by two vector order parameters. It is 
found that the AMR and PHR in such materials have universal angular dependences. 
In chiral magnets, for example materials with DMI, both longitudinal 
and transverse dc resistivity is predicted to be non-reciprocal.  
Different from polycrystalline magnetic materials with only one order parameter 
where AMR and PHR have the same magnitude, and $\pi/4$ out of phase, the magnitude 
of AMR and PHR of materials with two order parameters are not the same in 
general, and the phase difference is not $\pi/4$. Instead of $\pi$ periodicity 
of the usual AMR and PHR, the periodicities of AMR and PHR in magnetic materials 
with two order parameters are $2\pi$. 

\begin{acknowledgments}
This work is supported by the National Key Research and Development Program 
of China (No. 2020YFA0309600 and 2018YFB0407600), the NSFC Grant 
(No. 11974296 and 11774296) and Hong Kong RGC Grants (No. 16301518, 
16301619, and 16302321). 
\end{acknowledgments}


\begin{thebibliography}{99}

\bibitem{Kelvin}W. Thomson, ``On the Electro-Dynamic Qualities of 
Metals: — Effects of Magnetization on the Electric Conductivity of 
Nickel and of Iron", Proc. Royal Soc. Lond. 8: 546–550 (1857). 
doi:10.1098/rspl.1856.0144. 

\bibitem{book1} I. A. Campbell and A. Fert, {\it Transport 
Properties of Ferromagnets, in Handbook of Ferromagnetic Materials} 
(Elsevier, New York, 1982), Vol. 3, Chap. 9 pp. 747–804,
https://doi.org/10.1016/S1574-9304(05)80095-1.

\bibitem{book2}R. C. O’Handley, {\it Modern Magnetic Materials: 
Principles and Applications} (Wiley, New York, 2000).

\bibitem{CFJ}I. A. Campbell, A. Fert, O. Jaoul, ``The Spontaneous 
Resistivity Anisotropy in Ni-Based Alloys". J. Phys. C. 3, S95–S101 (1970). 

\bibitem{Tsunoda1}S. Kokado, M. Tsunoda, K. Harigaya and A. Sakuma,  
J. Phys. Soc. Jpn. 81, 024705 (2012).

\bibitem{Tsunoda2}S. Kokado, M. Tsunoda, ``Anisotropic Magnetoresistance 
Effect: General Expression of AMR Ratio and Intuitive Explanation for 
Sign of AMR Ratio", Advanced Materials Research 750-752, 978-982 (2013).

\bibitem{smit}J. Smit, Physica 17, 612 (1951).

\bibitem{McGuire1}T. R. McGuire and R. Potter,``Anisotropic magnetoresistance 
in ferromagnetic 3d alloys", IEEE Transactions on Magnetics 11, 1018 (1975).

\bibitem{McGuire2}T. R. McGuire, J. A. Aboafand, and E. Klokholm, 
IEEE Trans. Magn. 20, 972 (1984).

\bibitem{Ohandley}R. C. O’Handley, Phys. Rev. B 18, 2577 (1978).

\bibitem{Wisniewski}P. Wisniewski,``Giant anisotropic magnetoresistance 
and magnetothermopower in cubic 3:4 uranium pnictides", 
Applied Physics Letters. 90 (19): 192106 (2007). 

\bibitem{sin-cry-9}F. L. Zeng, Z. Y. Ren, Y. Li, J. Y. Zeng, M. W. Jia, J. Miao, A. Hoffmann, 
W. Zhang, Y. Z. Wu, and Z. Yuan, ''Intrinsic mechanism for anisotropic magnetoresistance and 
experimental confirmation in CoxFe1-x single-crystal films", Phys. Rev. Lett. 125, 
097201 (2020).

\bibitem{sin-cry-10}L. Nadvornik, M. Borchert, L. Brandt, R. Schlitz, K. A. de Mare, K. Vyborny,  
I. Mertig, G. Jakob, M. Klaui, S. T. B. Goennenwein, M. Wolf, G. Woltersdorf, and T. Kampfrath,
``Broadband Terahertz Probes of Anisotropic Magnetoresistance Disentangle Extrinsic and 
Intrinsic Contributions", Phys. Rev. X 11, 021030 (2021). 

\bibitem{gerrit}G. E. W. Bauer, ``Anisotropic magnetoresistance: A 
170-year-old puzzle solved", SCIENCE CHINA: Physics, Mechanics \& 
Astronomy 64, 217531 (2021).

\bibitem{Jungwirth}E. De Ranieri, A. W. Rushforth, K. Vyborny, U. Rana, E. Ahmad, 
R. P. Campion, C. T. Foxon, B. L. Gallagher, A. C. Irvine, J. Wunderlich, and 
T. Jungwirth, New J. Phys. 10, 065003 (2008).

\bibitem{yin1}Y. Zhang, H. W. Zhang, and X. R. Wang, ``Extraordinary 
galvanomagnetic effects in polycrystalline magnetic films", 
Europhys. Lett. 113, 47003 (2016).

\bibitem{sin-cry-1}M. Tondra, D. K. Lottis, K. T. Riggs, Y. Chen, E. D. Dahlberg, 
and G. A. Prinz, ``Thickness dependence of the anisotropic magnetoresistance in 
epitaxial iron films", J. Appl. Phys. 73, 6393 (1993).

\bibitem{sin-cry-2}R. P. van Gorkom, J. Caro, T. M. Klapwijk, and S. Radelaar, 
``Temperature and angular dependence of the anisotropic magnetoresistance in 
epitaxial Fe films", Phys. Rev. B 63, 134432 (2001).

\bibitem{sin-cry-3}W. Limmer, M. Glunk, J. Daeubler, T. Hummel, W. Schoch,
R. Sauer, C. Bihler, H. Huebl, M. S. Brandt, and S. T. B. Goennenwein, 
``Angle-dependent magnetotransport in cubic and tetragonal ferromagnets: 
Application to (001)- and (113)A-oriented (Ga,Mn)As", Phys. Rev. B 74, 205205 (2006).

\bibitem{sin-cry-4}W. Limmer, J. Daeubler, L. Dreher, M. Glunk, W. Schoch, S. Schwaiger, 
and R. Sauer, ``Advanced resistivity model for arbitrary magnetization orientation applied 
to a series of compressive- to tensile-strained (Ga,Mn)As layers", Phys. Rev. B 77, 205210 (2008).

\bibitem{sin-cry-5}Y. Bason, J. Hoffman, C. H. Ahn, and L. Klein, ``Magnetoresistance tensor of 
La0.8Sr0.2MnO3", Phys. Rev. B 79, 092406 (2009).

\bibitem{sin-cry-6}N. Naftalis, A. Kaplan, M. Schultz, C. A. F. Vaz, J. A. Moyer, 
C.H. Ahn, and L. Klein, ``Field-dependent anisotropic magnetoresistance and planar Hall 
effect in epitaxial magnetite thin films", Phys. Rev. B 84, 094441 (2011).

\bibitem{sin-cry-7}Z. Ding, J.X. Li, J. Zhu, T.P. Ma, C. Won, and Y.Z. Wu, 
``Three-dimensional mapping of the anisotropic magneto- resistance in Fe3O4 single 
crystal thin films", J. Appl. Phys. 113, 17B103 (2013).

\bibitem{sin-cry-8}T. Hupfauer, A. Matos-Abiague, M. Gmitra, F. Schiller, J. Loher, 
D. Bougeard, C. H. Back, J. Fabian, and D. Weiss, ``Emergence of spin-orbit fields in 
magnetotransport of quasi- two-dimensional iron on gallium arsenide", Nat. Commun. 6, 7374 (2015).

\bibitem{haitao}H. T. Wu, X. C. Hu, and X. R. Wang, ``Nematic and smectic stripe phases 
and stripe-SkX transformations", SCIENCE CHINA: Physics, Mechanics \& 
Astronomy http://engine.scichina.com/doi/10.1007/s11433-021-1852-8.

\bibitem{Sakurai}Section 3.10 in {\it Modern Quantum Mechanics}, 
J. J. Sakurai/San Fu Tuan, Revised Edition, Addison Wesley Longman, 1994.

\bibitem{yin2}Y. Zhang, X. S. Wang, H. Y. Yuan, S. S. Kang, H. W. Zhang, 
and X. R. Wang, ``Dynamic magnetic susceptibility and electrical detection of 
ferromagnetic resonance", \textit{J. Phys.: Condens.  Matter} \textbf{29}, 095806 (2017).

\bibitem{Carmeli}Sections 2.9 and 3.1 in {\it Classical Fields: General Relativity 
and Gauge Theory}, Moshe Carmeli, World Scientific Publishing Company, 2001.

\bibitem{xrw}X. R. Wang, ``Anomalous spin hall and inverse spin hall effects in magnetic 
systems", Commun. Phys. 4, 55 (2021).

\bibitem{yin3}Y. Zhang, Q. Liu, B. F. Miao, H. F. Ding, and X. R. Wang, 
``Anatomy of electrical signals and dc-voltage line shape in spin-torque 
ferromagnetic resonance", Phys. Rev. B \textbf{99}, 064424 (2019).

\end{thebibliography}
\end{document}